\documentclass[9pt,twocolumn,twoside]{osajnl}
\usepackage{graphicx}
\usepackage{subcaption}
\usepackage{bm}

\journal{ol} 

\setboolean{shortarticle}{true}

\title{Passively mode-locked thulium-doped all-fiber laser based on low V number fiber bending}

\author[1]{Hongbo Jiang}
\author[1]{Yunpeng Huang}
\author[1]{Zihao Zhao}
\author[1]{Lei Jin}
\author[1]{Shinji Yamashita}
\author[1, *]{Sze Y. Set}

\affil[1]{The University of Tokyo, Research Center for Advanced Science and Technology, 4-6-1 Komaba, Meguro-ku, Tokyo 153-8904, Japan}

\affil[*]{Corresponding author:
set@cntp.t.u-tokyo.ac.jp}

\begin{abstract}
This paper describes a mode locking technique based on enhanced polarization dependent loss (PDL). The method utilizes a bent single mode fiber (SMF28) coil to induce sufficient PDL at 2 \(\mu\)m wavelength region. Significant PDL in SMF28 coils is enabled since the light is much more weakly guided. A passively mode-locked thulium doped all-fiber laser is demonstrated using this simple device with polarization controllers as mode locker. The results indicate a moderate amount of 1 dB is sufficient to initiate and sustain stable mode-locking operation. We believe, to the best of our knowledge, this is the first demonstration of mode-locked fiber laser fully based on polarization dependent property of bent fiber section. 
\end{abstract}

\setboolean{displaycopyright}{true}

\begin{document}

\maketitle

Passively mode-locked fiber lasers have been widely used for decades due to their simplicity and their ability to generate ultra-short optical pulses \cite{Nelson1997,cite-key}. The key element in a passively mode-locked fiber laser is saturable absorber (SA), which has an intensity-dependent response to support optical pulse formation and against continuous-wave (CW) lasing \cite{902165,weiner2011ultrafast}. Various kinds of SAs have been demonstrated, such as semiconductor saturable absorber mirror (SESAM) \cite{571743}, single-wall carbon nanotube (SWNT) \cite{1266677} and graphene \cite{doi:10.1021/nn901703e}. Besides, there are also artificial SAs, whereby an effective saturable absorption property can be realized with configurations like nonlinear polarization evolution (NPE) \cite{256032,173059}, nonlinear optical loop mirror (NOLM) \cite{Doran:88,272080} and nonlinear amplifying loop mirror (NALM) \cite{Fermann:90}. All the above mentioned SAs have been studied in detail and many of them are commonly used in standard passively mode-locked fiber lasers.

The NPE mode locking method is superior in many aspects, such as a high optical power tolerance, a femtosecond-scale response time, and an all-fiber structure. In conventional NPE mode-locked fiber lasers, an optical polarizer or a polarization-sensitive isolator with the PDL of > 30 dB is employed. By a proper adjustment of the polarization controller to favor the transmission of higher intensity light through the polarizer, mode locking can be achieved to generate ultra-short pulses. In fact, NPE mode locking method can be realized without the polarizer/polarization-dependent isolator installed, as long as the intensity-dependent transmission effect in fiber laser enabled by PDL is sufficiently large, and the source of this PDL could be varied. In fact, numerous studies on PDL for NPE mode locking have been reported in recent years. An all-fiber NPE mode-locked fiber laser has been demonstrated using a 45-degree tilted grating as an in-fiber polarizer with PDL of 30 dB \cite{Zhang:13}. However, the fabrication of the 45-degree tilted fiber grating with such performance is not trivial. Wu \emph{et al} \cite{Wu2010} demonstrated a erbium-doped mode-locked fiber laser using a defective optical component with a residual PDL of 3 dB as the polarizing element, indicating that instead of using an optical component with large PDL (30 dB), a relatively small PDL (3 dB) in fiber laser is sufficient for NPE mode locking operation. Moreover, it has also been demonstrated that weak PDL caused by bent fiber in the laser cavity could initiate NPE mode locking. However, only noise-like pulses was demonstrated, stable fundamental CW mode locking operation was not possible \cite{Lin_2014}. It is likely due to the fact that the fiber bent induced PDL is not sufficiently significant for NPE mode locking. Since SMF28 is designed for operation at 1.55 \(\mu\)m, fiber bent induced PDL is relatively weak (estimated in the order of 0.1 dB \cite{Wang:07}). This could explain why only the pulses generation was reported, whereas the PDL was not measured or quantified in the experiment straightforwardly. 

Here, we propose a method to enhance the bend-induced PDL inside the single mode fiber. Instead of bending a length of fiber at its designed wavelength regime, we let the longer wavelength light pass through the bent fiber section. At a certain operating wavelength, due to the different degrees of optical intensity leakage, a section of the bent fiber with a lower normalized frequency (V number) will experience a larger PDL than bent fiber with a normal V number. In our case, at 2\(\mu\)m wavelength regime, SMF28 can be defined as lower V number (V=1.75) fiber (abbreviated as LVF for convenience). And correspondingly, the normal V number fiber (NVF) designed for 2\(\mu\)m such as SM1950 has the V number equals to 2.27. When properly coiling the LVF and carefully choosing the bending diameter, the PDL could be significantly increased up to 3.3 dB. And it is unlikely to get the equivalent amount of PDL using NVF, no matter how hard it is bent. Therefore, bent LVF coil could provide sufficient PDL compare to NVF, so that a new NPE mode locking approach becomes feasible. As a prove of concept, we construct a 2\(\mu\)m thulium-doped fiber laser system using a bent section of LVF as PDL element. When a bent LVF section with moderate PDL of 1 dB is installed, stable CW fundamental mode locking is successfully initiated and sustained. The results confirm the potential of bent LVF for laser mode-locking applications. Since the coiling of LVF is simple and reproducible, this technique preserves all the advantages of NPE mode locking while features low cost and all-fiber configuration at the same time. 

\begin{figure}[htbp]
\centering
\includegraphics[width=0.9\linewidth]{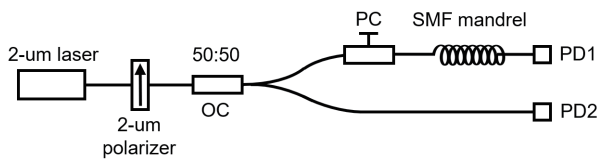}
\caption{Experimental setup for measuring the PDL of fiber coils. OC: optical coupler; PC: polarization controller; PD: photon diode.}
\label{fig:PDL-1}
\end{figure}

The bend loss and the PDL of LVF (SMF28) and NVF (SM1950) are measured with a 2 \(\mu\)m laser source. A 2-meter-long SMF28 patch cord and a 2-meter-long SM1950 patchcord are carefully coiled around different cylindrical rod. The bending diameter can be altered by simply changing the diameters of the cylindrical rod. Fig. \ref{fig:PDL-1} shows the experimental setup used for measuring the bend loss and the PDL based on polarization scanning method. In order to avoid the power fluctuation caused by the laser source to affect the results, a 3 dB optical fused coupler is used to separate the optical path into a test path and a reference path. The rest of the system consists of a self-made CW thulium-doped fiber laser, a 2 \(\mu\)m in-line polarizer, a polarization controller (PC) (OZ Optics, FPC100), a fiber coil under test, and two identical optical power sensors (Thorlabs, S148C). The minimum ($BL_{min}$) and maximum bend loss ($BL_{max}$) can be detected after traversing all possible polarization states via tuning the PC, and PDL property can be determined by $BL_{max}-BL_{min}$. 

\begin{figure}[t]
\centering  
    \begin{subfigure}[b]{0.495\columnwidth}
        \includegraphics[width=\columnwidth]{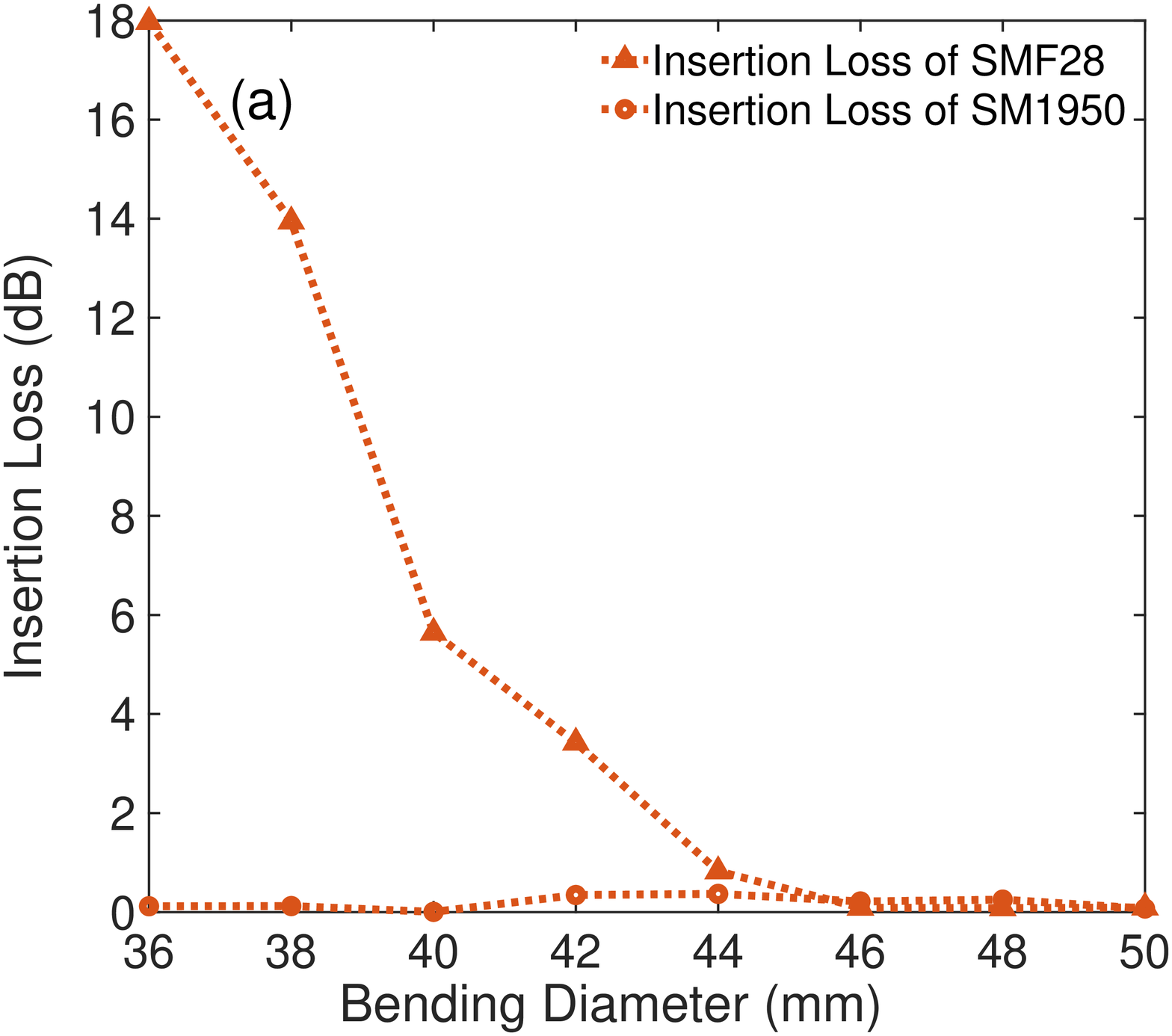}
        \label{fig:PDL2-a}
        \end{subfigure}
        \begin{subfigure}[b]{0.495\columnwidth}
        \includegraphics[width=\columnwidth]{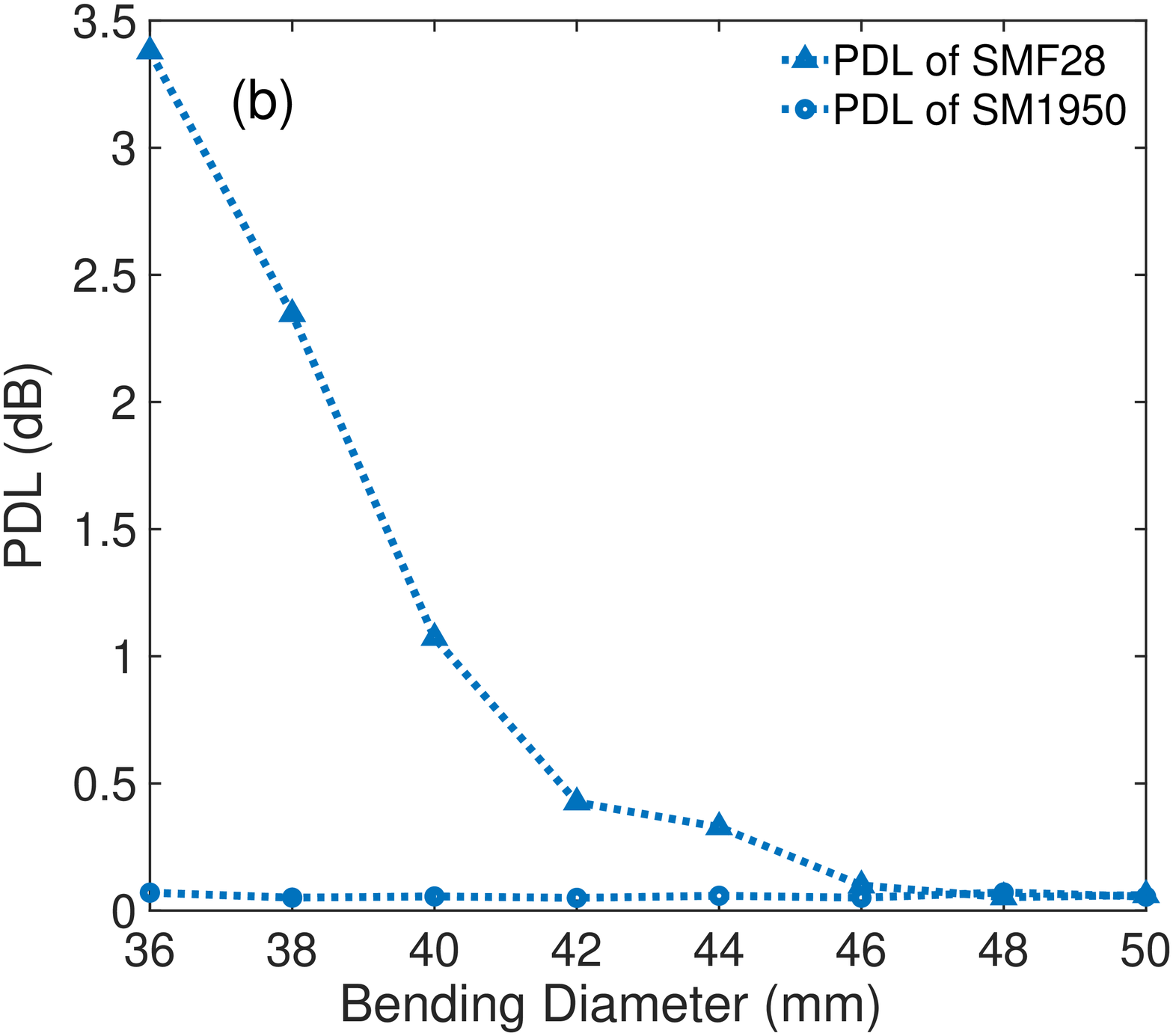}
        \label{fig:PDL2-b}
        \end{subfigure}
    \begin{subfigure}{1.0\columnwidth}
        \centering 
        \includegraphics[width=\columnwidth]{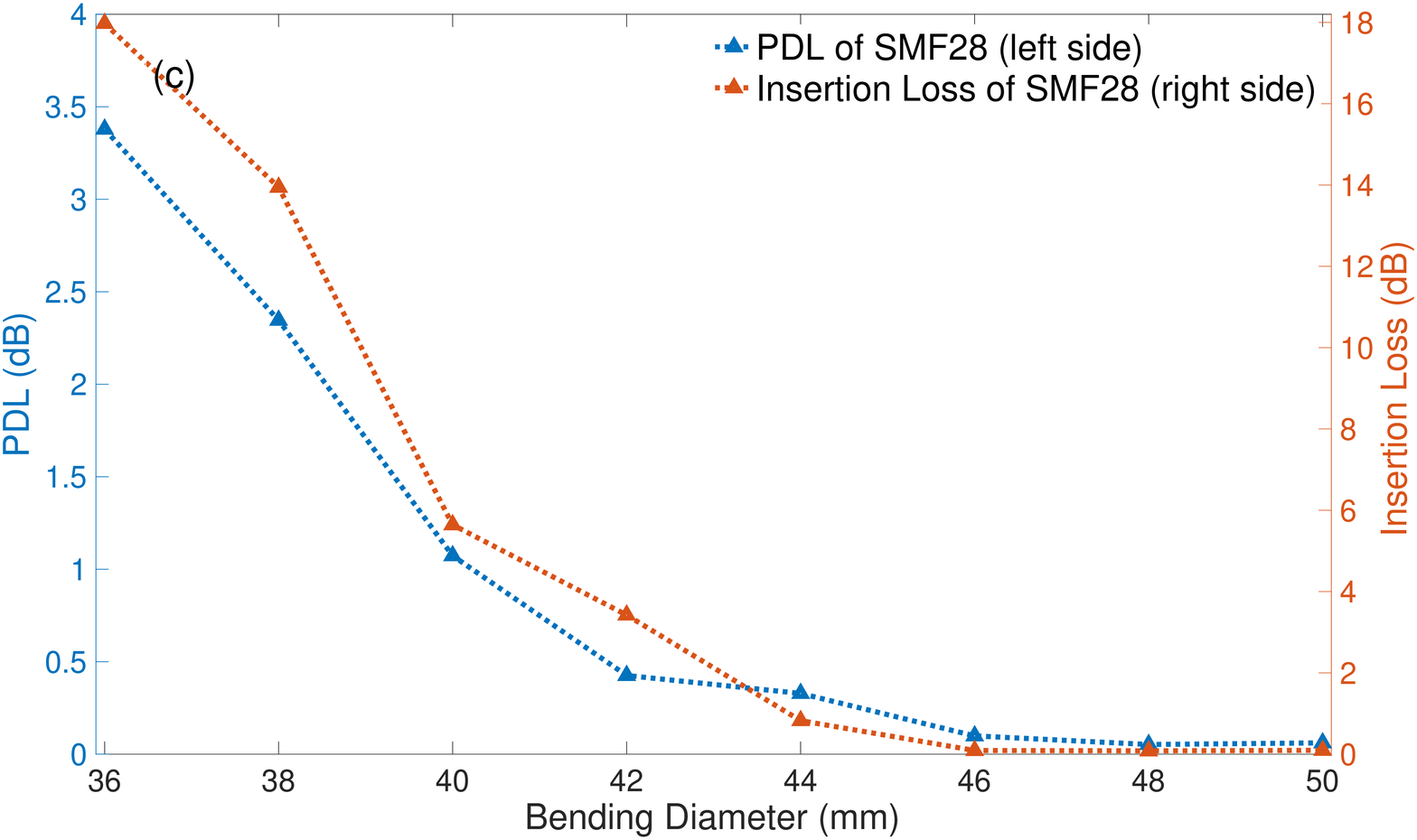}
        \label{fig:PDL2-c}      
    \end{subfigure}
\caption{Insertion loss comparison result between SMF28 and SM1950 fiber coils with different bending diameter (a), PDL comparison result  with different bending diameter (b), and results of SMF28 (c), respectively.} 
\label{fig:PDL-2}
\end{figure}

Fig. \ref{fig:PDL-2} shows the bend loss and the PDL characteristics of LVF coil and NVE coil with ten turns ($n=10$), within the bending diameter $d$ ranges from 36 mm to 50 mm. As shown in Fig. \ref{fig:PDL-2}a, the bend loss of LVF coil (indicated as red triangle) increases monotonically with decreasing $d$, and become significant (> 5 dB) when $d$ is < 40 mm. A maximum of more than 18 dB bend loss is recorded at $d$ = 36 mm. In contrast, NVF coil experiences a negligible bend loss (indicated as red dot) in the entire bending range. Since 2 \(\mu\)m light is much more weakly guided in LVF than in NVF, LVF is more sensitive to bending. As shown in Fig. \ref{fig:PDL-2}b, when the bending diameter decreases to $d$ < 40 mm, the amount of PDL in LVF coil (indicated as blue triangle) exceeds 1 dB. PDL even becomes more significant than 3 dB with further coiling the LVF to $d$ =36 mm. But PDL in NVF (indicated as blue dot) is still negligible in the entire bending range. The comparison results show that PDL generated in LVF is much more significant than it in NVF. Therefore, LVF is more suitable for accumulating PDL inside fiber. For the convenience of comparison, the PDL and bend loss results of LVF are depicted in Fig. \ref{fig:PDL-2}c. In sets of data with different $d$, although the PDL in LVF when $d$ = 36 mm and $d$ = 38 mm are significant, the excessive bend loss inside the fiber prevents the practical application. For large diameter like $d$ = 42 mm, although low bend loss is achieved, PDL of merely 0.4 dB might not be sufficient for NPE mode locking. It is worth to note that when $d$ = 40 mm, relatively moderate amount PDL of 1 dB along with an acceptable bend loss of 5.6 dB is achieved. We believe this LVF coil, or other coils with properties similar to this coil, could become suitable candidates for mode locking based on fiber bending. 

\begin{figure}[htbp]
\centering
    \begin{subfigure}[b]{0.495\columnwidth}
        \includegraphics[width=\columnwidth]{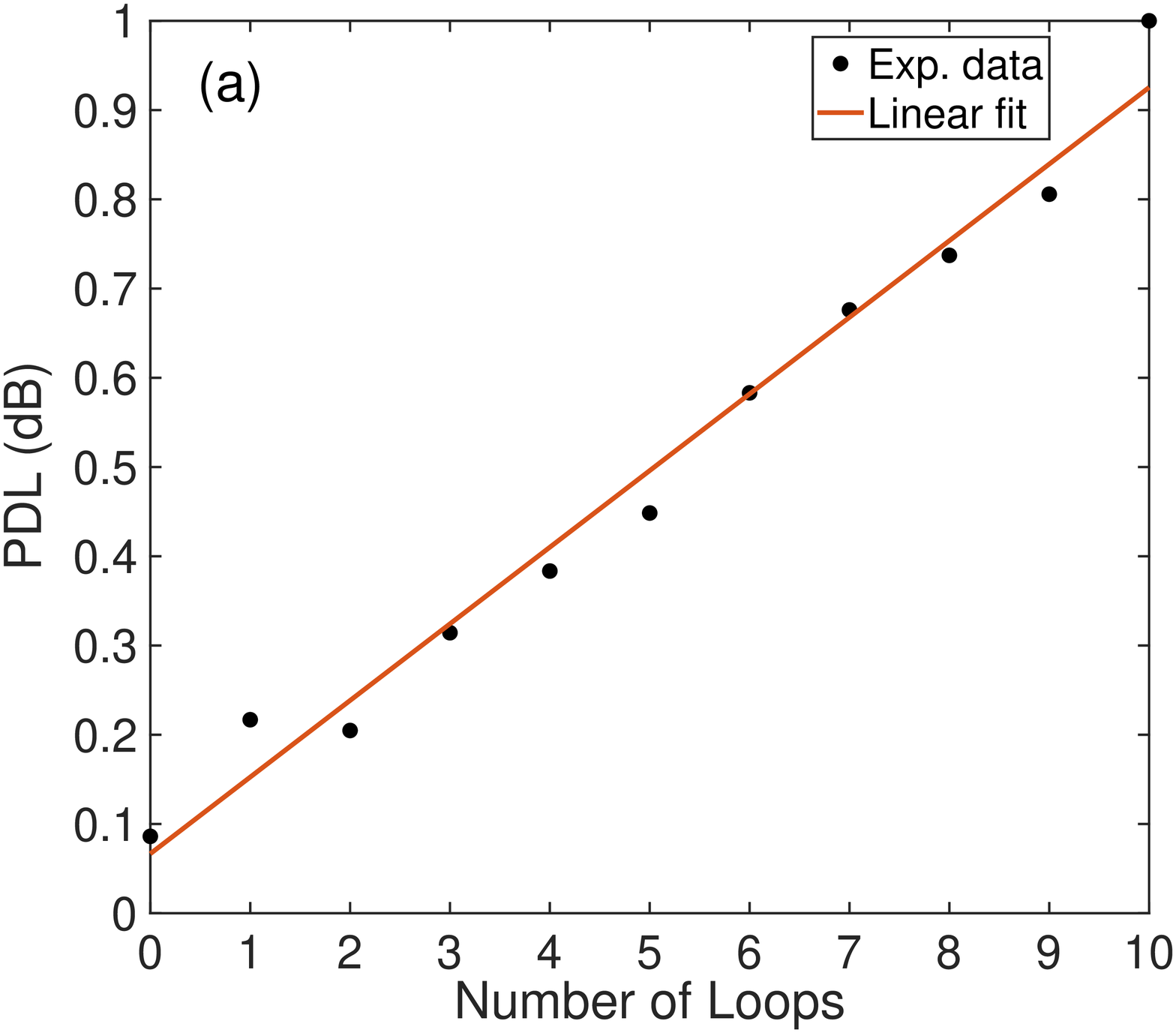}
        \label{fig:PDLvsLoops}
        \end{subfigure}
        \begin{subfigure}[b]{0.495\columnwidth}
        \includegraphics[width=\columnwidth]{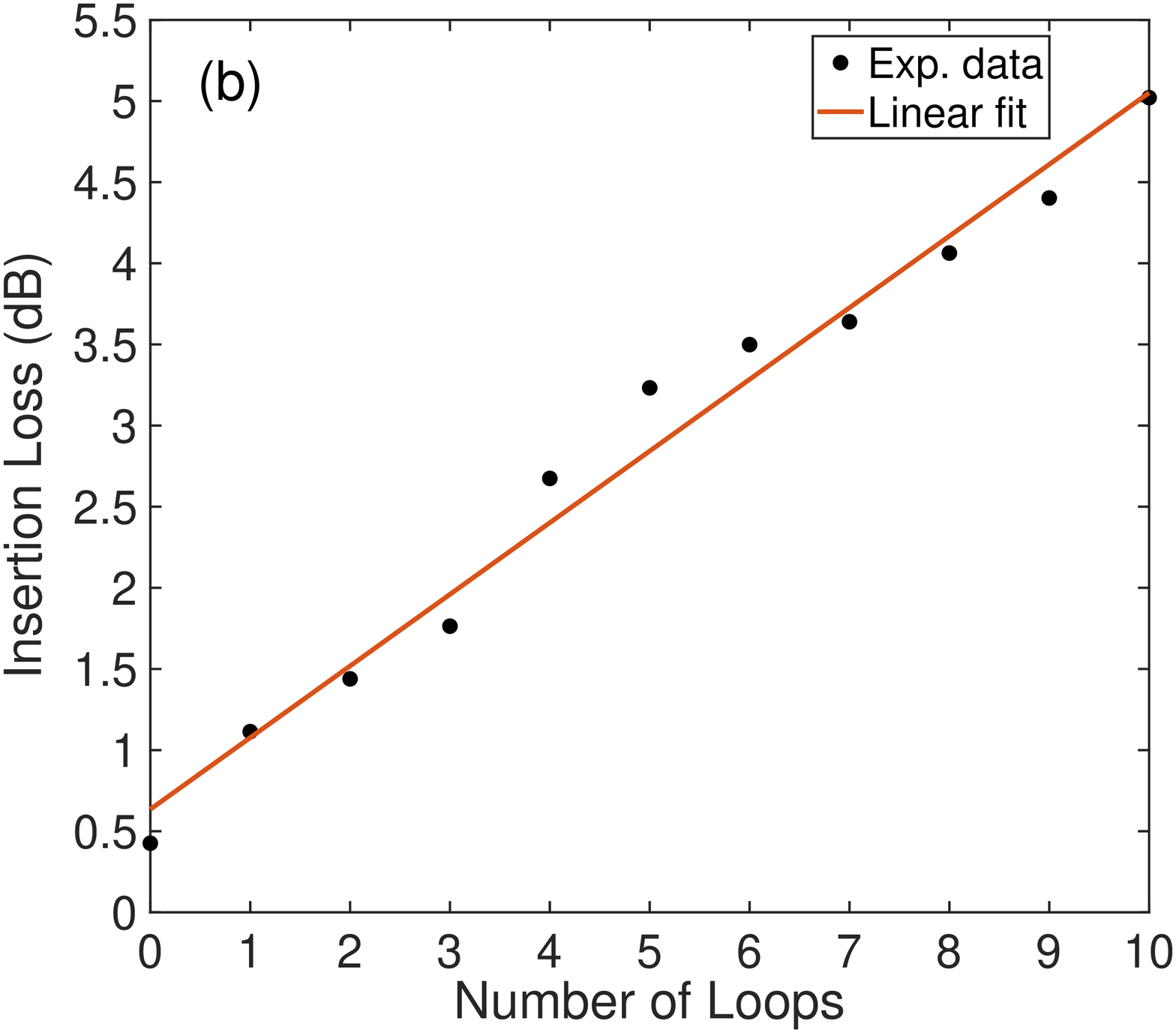}
        \label{fig:ILvsLoops}
        \end{subfigure}
\caption{Relationship between number of turns of the fiber coils with PDL (a) and bend loss (b), respectively.}
\label{fig:PDL-3}
\end{figure}

Moreover, the relationship between the number of coiling turns with the PDL and the bend loss in LVF are presented in Fig. \ref{fig:PDL-3}. Both the PDL and the bend loss increase monotonically with the turns number of the coil. The results match quite well with the linear fitting relationship, indicating that the PDL and bend loss properties are equally accumulated in each single loop in the LVF coil, and both of them exhibit roughly linear growth with number of turns. Besides, the residual effect at n=0 could be explained as imperfect splicing and un-straight fiber placement during the test.

\begin{figure}[htbp]
\centering
\includegraphics[width=0.95\linewidth]{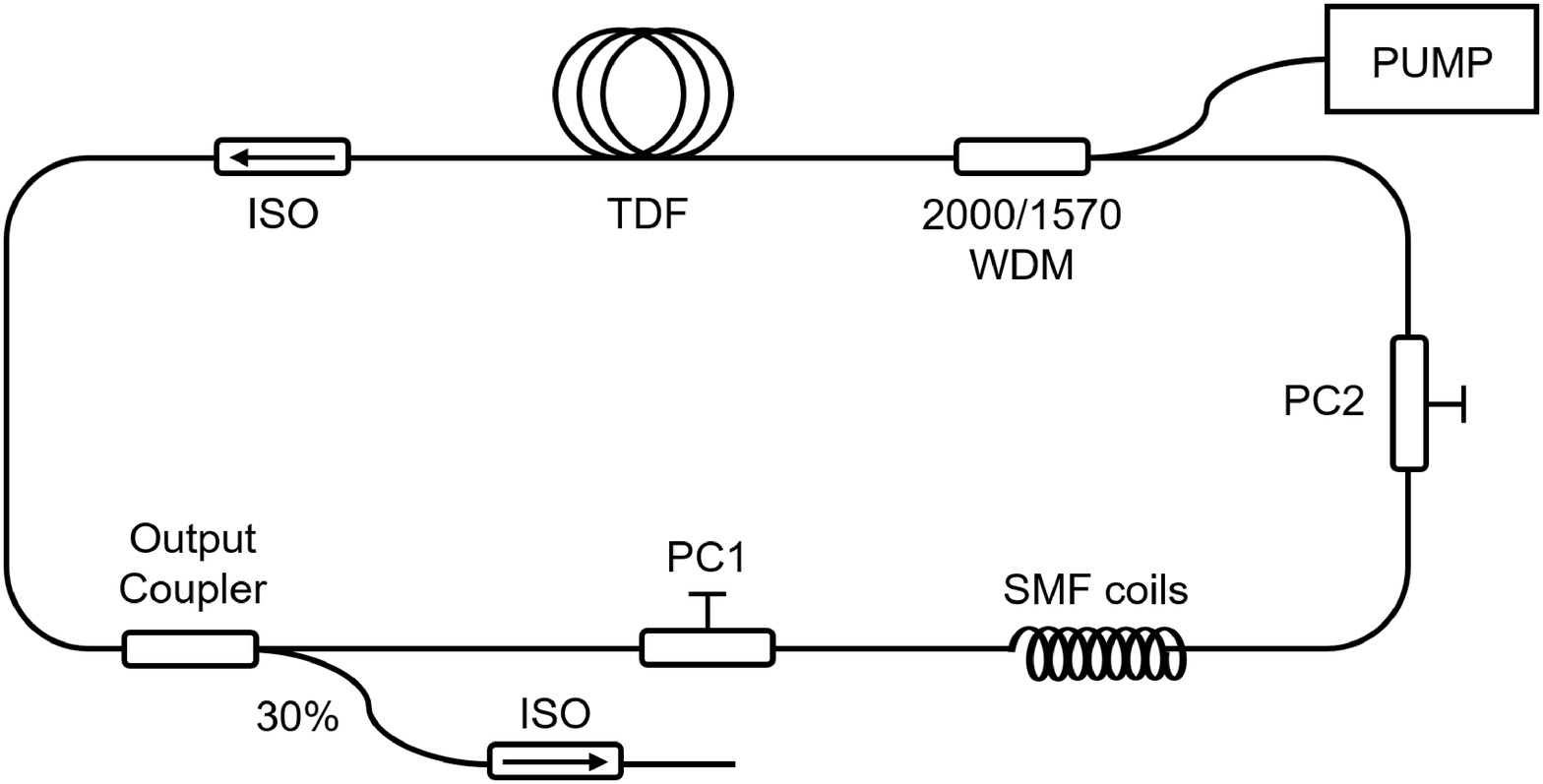}
\caption{Experimental setup of the thulium-doped mode-locked fiber ring laser using a fiber coils. ISO: isolator; TDF: thulium-doped fiber; WDM: wavelength division multiplexer; PC: polarization controller.}
\label{fig:laser-1}
\end{figure}

As shown in Fig. \ref{fig:laser-1}, LVF coil is installed in a thulium-doped fiber ring laser for mode locking test. An external cavity laser (Agilent 8168A) at 1570 nm amplified by an erbium-doped fiber amplifier (EDFA) is used as the pump source. A 3-meter-long of thulium-doped fiber (TDF) (OFS, TmDF200) is used as the laser gain medium. The gain medium and the pump source are connected to the cavity through a 1570/2000 nm wavelength-division-multiplexing (WDM) fused coupler. An optical isolator (ISO) is inserted to prevent backreflection in the cavity and to ensure unidirectional operation. The output of the laser is tapped through a 30\% port of a fiber fused coupler, whereas the other 70\% port is looped back to the cavity. A in-line-typed polarization controller (OZ Optics, FPC-100) consists with a segment of bare NVF is used to transform the linearly polarized light into elliptically polarized light. A second PC is inserted after the LVF coil to re-linearize the polarization state inside the cavity. The pigtails of all the passive fiber components are made of NVF, to ensure that cavity has minimal bending sensitivity. So the impact of PDL is solely originated from the LVF coil. The coil with PDL of 1 dB and bend loss of 5.6 dB ($d$=40 mm, $n$=10) is tested at first due to its moderate PDL and the acceptable bend loss. The total cavity length is about 20 m.

\begin{figure}[htbp]
\centering
    \begin{subfigure}[b]{0.495\columnwidth}
        \includegraphics[width=\columnwidth]{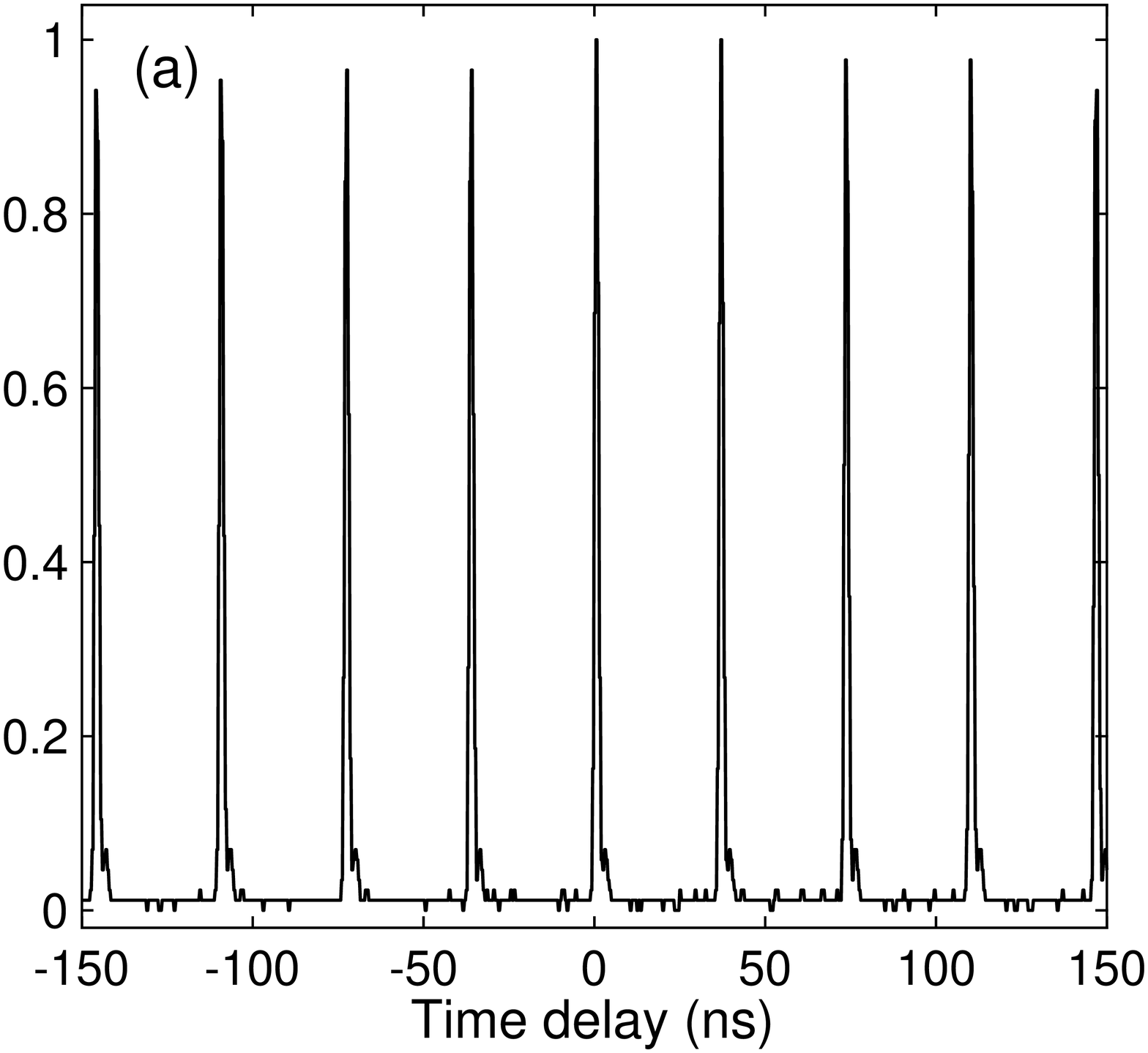}
        \label{fig:ms}
        \end{subfigure}
        \begin{subfigure}[b]{0.495\columnwidth}
        \includegraphics[width=\columnwidth]{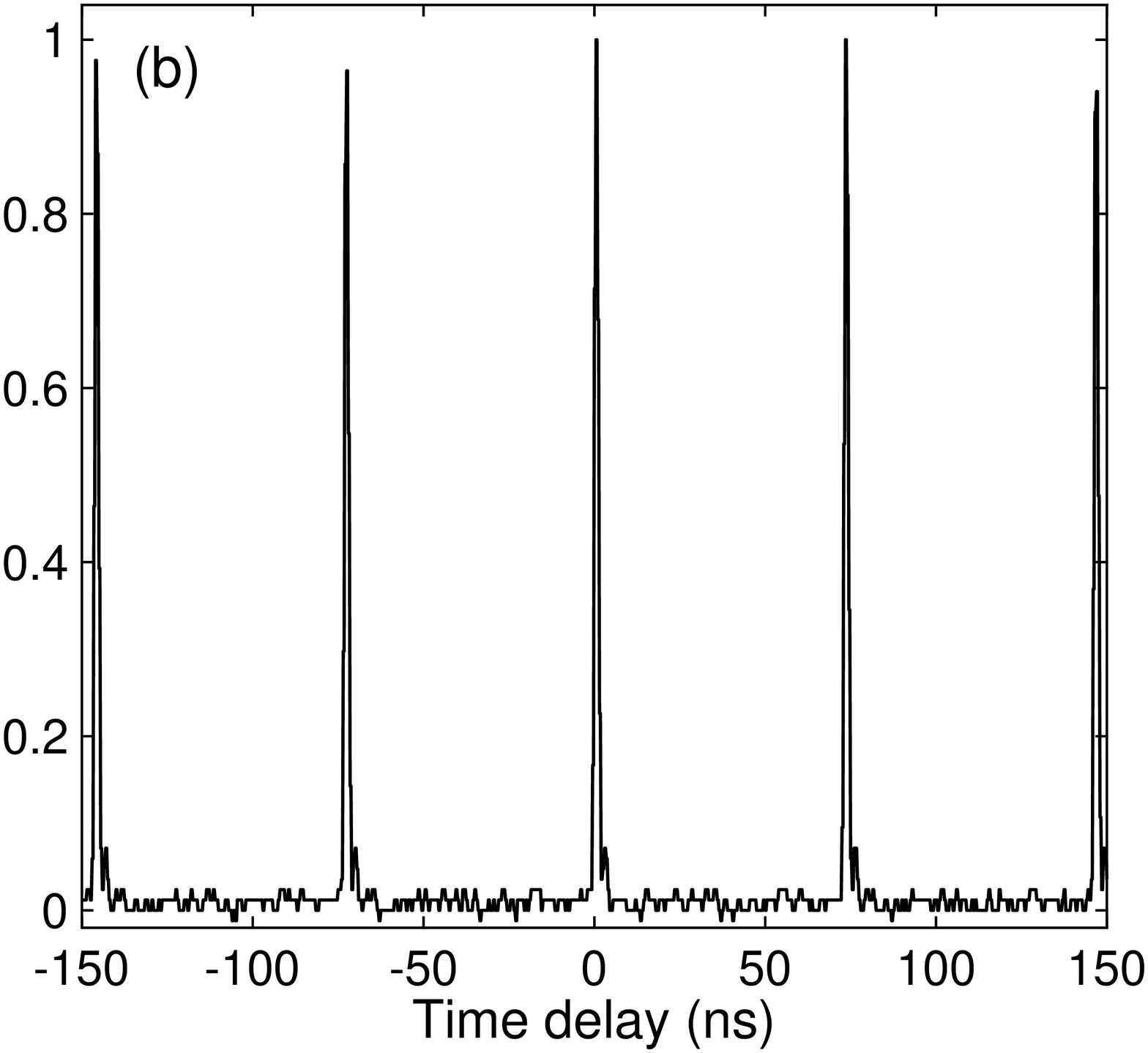}
        \label{fig:ss}
        \end{subfigure}
\caption{The waveform of the output pulse from the laser operating in harmonic mode (a) and fundamental mode (b).}
\label{fig:ms and ss}
\end{figure}

At a relatively high pump power about 200 mW, the laser starts mode locking operation in the multiple-pulse mode. The laser can be operated in a single-pulse mode when the pump power is lowered down to 170 mW. Output waveform of multiple-pulse mode and fundamental mode are shown in Fig. \ref{fig:ms and ss}.

\begin{figure}[htbp]
\centering
\includegraphics[width=1\linewidth]{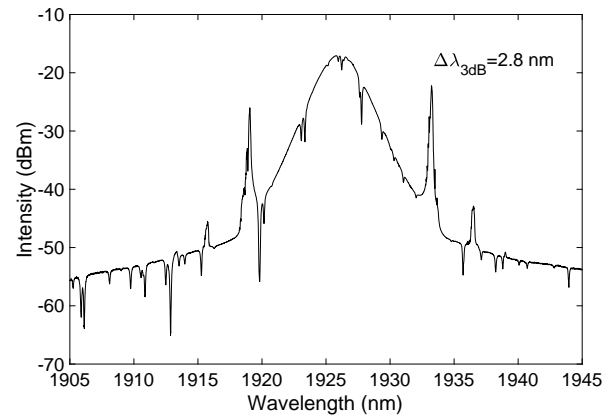}
\caption{Output spectrum of the mode-locked laser.}
\label{fig:osa}
\end{figure}

\begin{figure}[htbp]
\centering
    \begin{subfigure}[b]{0.495\columnwidth}
        \includegraphics[width=\columnwidth]{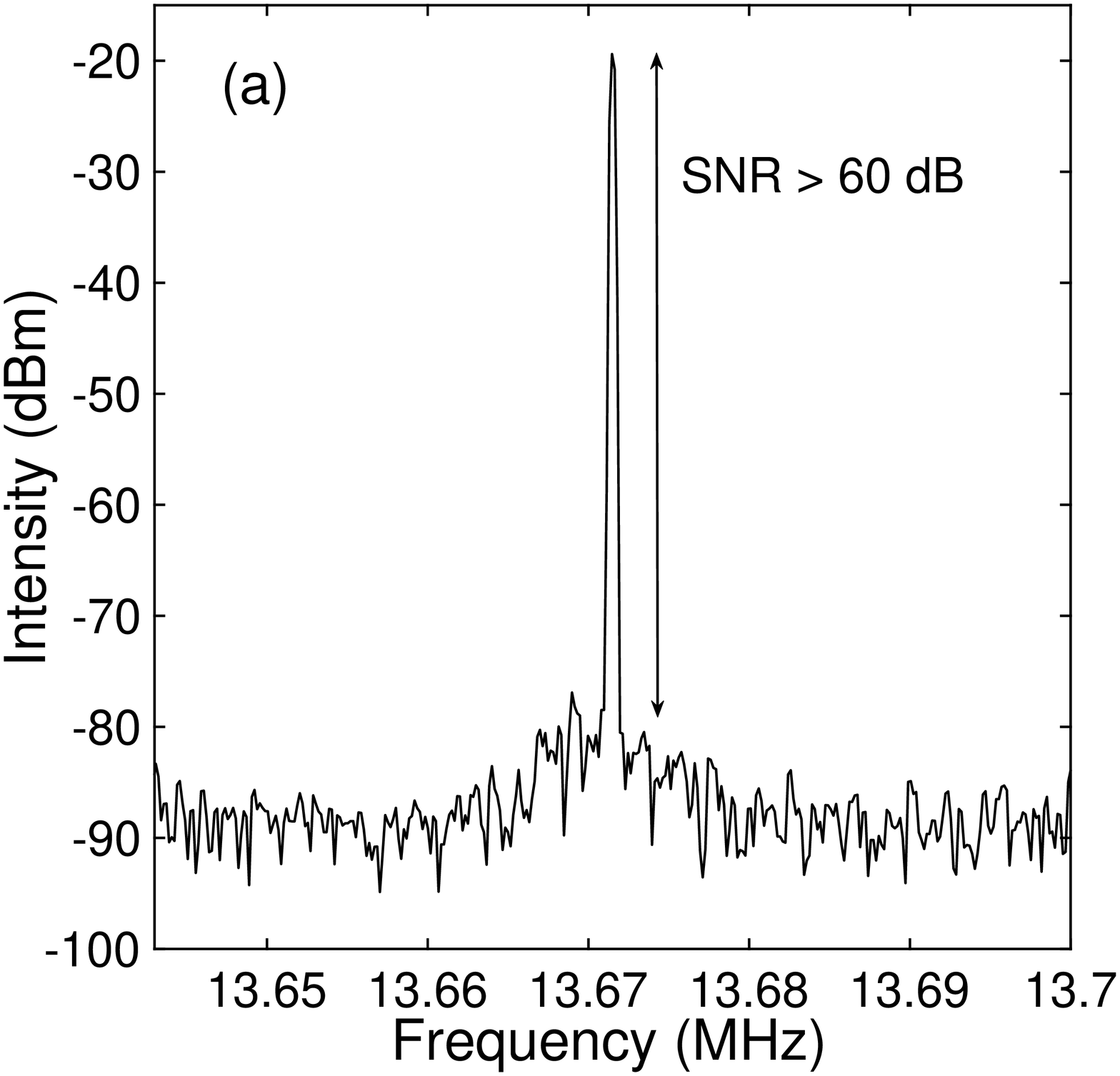}
        \label{fig:RF}
        \end{subfigure}
        \begin{subfigure}[b]{0.495\columnwidth}
        \includegraphics[width=\columnwidth]{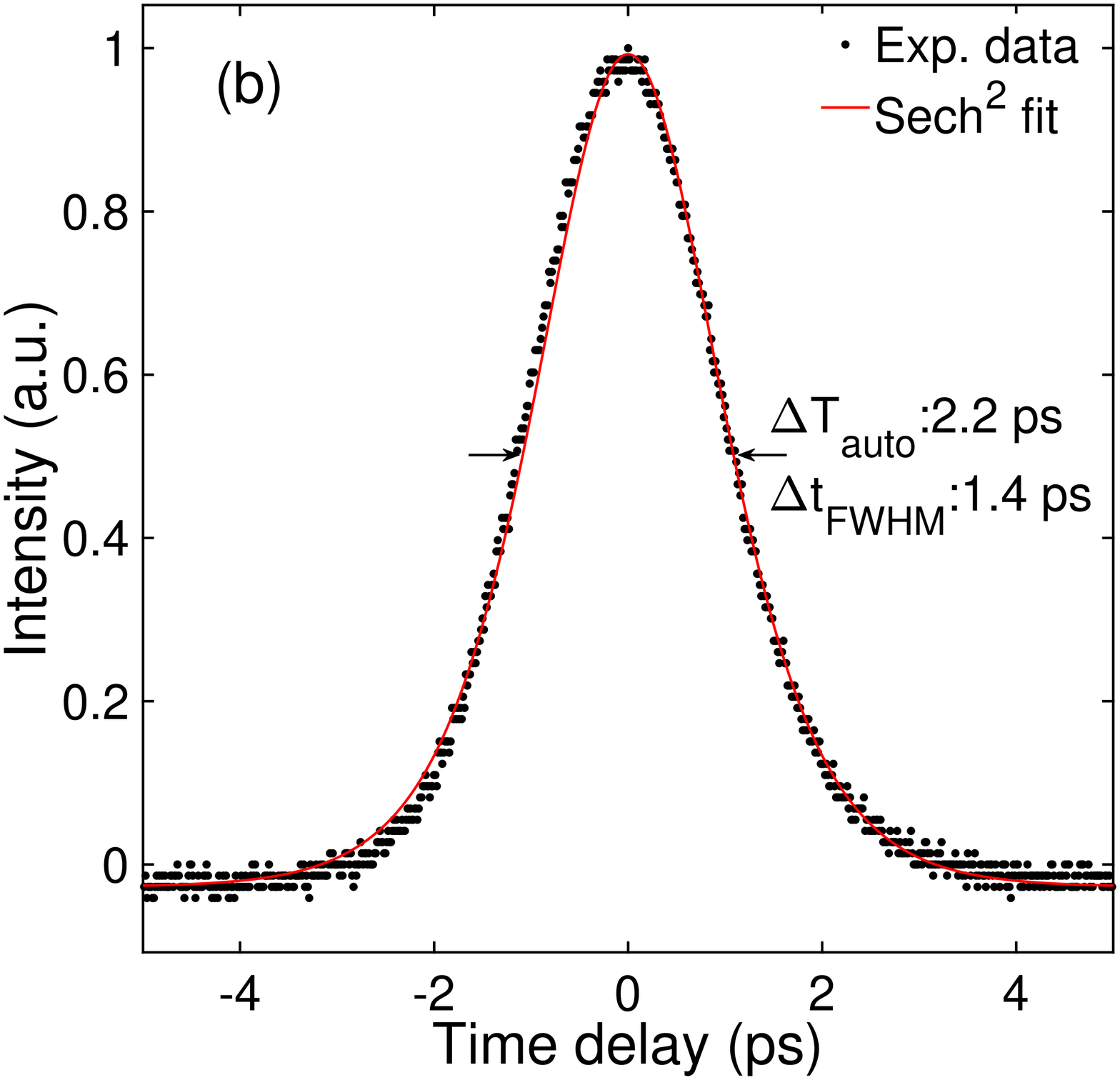}
        \label{fig:AC}
        \end{subfigure}
\caption{The RF spectrum (a) and the autocorrelation trace of the fundamental soliton (b).}
\label{fig:RF and AC}
\end{figure}

The output spectrum of the laser is shown in Fig. \ref{fig:osa}. The spectrum is measured with an optical spectrum analyser (YOKOGAWA, AQ6375) with an optical resolution bandwidth of 0.05 nm. The full width at half-maximum (FWHM) bandwidth is 2.8 nm with central wavelength at 1925 nm. The fundamental repetition rate is measured by using a photodetector with 10GHz bandwidth (EOT, ET-5000), connected with an RF spectrum analyzer (RIGOL, DSA832) with 100 kHz span and 100 Hz resolution bandwidth. As shown in Fig. \ref{fig:RF and AC}, the fundamental repetition rate is 13.67 MHz, with a radio-frequency signal-to-noise ratio of 60 dB. The autocorelation trace of the output pulse is measured with an autocorrelator (FEMTOCHROME, FR-103HP), well fitted by a sech pulse profile. The inferred FWHM width is estimated to be 1.4 ps, resulting in a time-bandwidth product (TBP) of 0.32, which is close to 0.315, the TBP of a trnsform-limited soliton pulse. The pulse average power is 4.7 mW, the pulse energy of 0.34 nJ is calculated. The corresponding peak power is 245 W.

When the LVF coil ($d$=40 mm, $n$=10) is removed from the laser cavity, or replaced with any other coils, it is not possible to mode lock the laser even when the pump power is raised to 350 mW. 

In the previous demonstration on bent SMF at 1.55\(\mu\)m, the PDL induced by fiber bending is very weak. In order to increase the value of PDL, bent section has to be coiled in smaller diameter, resulting in large bend losses. For instance, if the PDL value of 1 dB is expected by bending SMF28, the bend loss of 30 dB is inevitably introduced in the coil. It is hard to make full use of bending induced PDL with such huge intrinsic bend loss. Therefore, when fiber is bent at its designed wavelength, either insufficient PDL or excessive bend loss has hindered its potential for mode locking application. However, when SMF28 is used as LVF at 2\(\mu\)m wavelength regime, significant amount of PDL could be created. Since lower V number fiber is more sensitive to macrobending, LVF exhibits more PDL by leaking more light from core to cladding when it is bent. As a result, PDL accumulated in LVF coil is more significantly enhanced than the bend loss. In other words, when the same quantity of PDL is introduced, LVF exhibits much smaller bend loss than NVF. For instance, when the PDL value of 1 dB is obtained in LVF coil, the corresponding bend loss is 5.6 dB. Ideally, an element with higher PDL and lower insertion loss (bend loss in our case) is more effective for NPE mode locking. Here, a figure of merit (FOM) is defined to represent the suitability of different fiber coils for mode locking. 
\begin{equation}
FOM = {P}^2{T}*1000
\label{eq:FOM}
\end{equation}
where $P$ is a measure of polarizing efficiencies and $T$ is the transmissions of coils, respectively. The result is multiplied by 1000 to make the difference more evident. Polarizing efficiency $P$ is the power ratio converted from PDL, while transmittance $T$ is converted from bend loss. As shown in Table. \ref{tab:FOM}, the maximum value of FOM appears on the LVF coil with $d$=40 mm and $n$=10, which successfully initiates and sustains the stable fundamental mode locking operation. In addition, FOM of other coils are very small due to either insufficient polarizing efficiency or small transmittance. And correspondingly, the mode locking tests of these coils have all failed. Moreover, in order to highlight the LVF concept we have proposed. Here, the FOM value of NVF coil using experimental data from \cite{Wang:07} is calculated as reference. There is a noticeable gap between the reference and maximum result, which reflects the in-applicability of NVF caused by excessive loss and insufficient PDL. This could be a possiable reason to explain why it is hard to make full use of SMF28 bending at 1.55\(\mu\)m in mode locking application. Therefore, in terms of being used as PDL element in fiber laser, the usage of LVF is more suitable, in comparison with NVF. 
\begin{table}[htbp]
\centering
\begin{tabular}{cccccc}
\hline
Coils&PDL(dB)&$\bm{P}(\%)$&Bend Loss(dB)&$\bm{T}(\%)$&FOM\\
\hline
$d=36 mm$ & 3.38 & 54.07  &  18.01 & 1.58 & 4.62 \\
$d=38 mm$ & 2.34 & 41.68  &  16.50 & 2.24 & 3.89 \\
$d=40 mm$ & 1.07 & 21.89  &  5.68 & 27.05 & 12.96 \\
$d=42 mm$ & 0.43 & 9.33  &  3.46 & 45.10 & 3.93 \\
$d=44 mm$ & 0.33 & 7.28  &  0.86 & 82.03 & 4.35 \\
NVF(\cite{Wang:07})& 0.85 & 17.78  & 12 & 6.31 & 1.99 \\
\hline
\end{tabular}
\caption{FOM of various fiber coils}
\label{tab:FOM}
\end{table}

Since the bend loss in LVF coil could be simply compensated with extra gain in the cavity, the bent fiber section can be used as a mode locker in a very simple way. We believe, by using bent HI1060 in erbium-doped fiber lasers and SM780 in ytterbium-doped fiber lasers, it is also possible to achieve stable CW NPE mode locking operation. Compared with other saturable absorbers for passively mode locking, the bent fiber section is superior in the aspect of preparation process and the cost. On the other hand, the study unveils the impact of PDL in passively mode-locked fiber laser. Even if a little amount of PDL could be functionalized as an optical polarizer and profoundly affects the output. Sometimes the NPE effect enabled by PDL should be prevented in passively mode-locked fiber laser. In a non-polarization maintaining (PM) fiber laser system which commonly has paddle-typed PCs and bunch of fiber coils, extra care should be taken into the study of mode locking mechanism. Therefore, it is not always a good sign that only 1 dB polarization dependent discrimination is sufficient to initiate mode locking. Therefore, when this potentially misleading NPE need to be completely excluded in some cases, PM fiber laser is recommended. 

In conclusion, we have demonstrated a mode locking method based on bend-induced PDL. It is found that the PDL is significantly enhanced in single mode fiber with lower V number. A thulium-doped fiber laser is constructed with a section of SMF28 coil installed. Stable fundamental soliton generation is enabled by the LVF with PDL of 1 dB. Since this method only needs single mode fiber and simple coiling process, it has superior advantages in mode locking application such as all-fiber configuration and extreme cost-effectiveness.










\bibliography{sample}

\begin{thebibliography}{10}
\newcommand{\enquote}[1]{``#1''}

\bibitem{Nelson1997}
L.~Nelson, D.~Jones, K.~Tamura, H.~Haus, and E.~Ippen,
  \enquote{Ultrashort-pulse fiber ring lasers,} {\protect\JournalTitle{Applied
  Physics B}} \textbf{65}, 277--294 (1997).

\bibitem{cite-key}
M.~E. Fermann and I.~Hartl, \enquote{Ultrafast fibre lasers,}
  {\protect\JournalTitle{Nature Photonics}} \textbf{7}, 868 EP -- (2013).

\bibitem{902165}
H.~A. {Haus}, \enquote{Mode-locking of lasers,} {\protect\JournalTitle{IEEE
  Journal of Selected Topics in Quantum Electronics}} \textbf{6}, 1173--1185
  (2000).

\bibitem{weiner2011ultrafast}
A.~Weiner, \emph{Ultrafast optics}, vol.~72 (John Wiley \& Sons, 2011).

\bibitem{571743}
U.~{Keller}, K.~J. {Weingarten}, F.~X. {Kartner}, D.~{Kopf}, B.~{Braun}, I.~D.
  {Jung}, R.~{Fluck}, C.~{Honninger}, N.~{Matuschek}, and J.~{Aus der Au},
  \enquote{Semiconductor saturable absorber mirrors (sesam's) for femtosecond
  to nanosecond pulse generation in solid-state lasers,}
  {\protect\JournalTitle{IEEE Journal of Selected Topics in Quantum
  Electronics}} \textbf{2}, 435--453 (1996).

\bibitem{1266677}
S.~Y. {Set}, H.~{Yaguchi}, Y.~{Tanaka}, and M.~{Jablonski}, \enquote{Laser mode
  locking using a saturable absorber incorporating carbon nanotubes,}
  {\protect\JournalTitle{Journal of Lightwave Technology}} \textbf{22}, 51--56
  (2004).

\bibitem{doi:10.1021/nn901703e}
Z.~Sun, T.~Hasan, F.~Torrisi, D.~Popa, G.~Privitera, F.~Wang, F.~Bonaccorso,
  D.~M. Basko, and A.~C. Ferrari, \enquote{Graphene mode-locked ultrafast
  laser,} {\protect\JournalTitle{ACS Nano}} \textbf{4}, 803--810 (2010). PMID:
  20099874.

\bibitem{256032}
V.~J. {Matsas}, T.~P. {Newson}, D.~J. {Richardson}, and D.~N. {Payne},
  \enquote{Selfstarting passively mode-locked fibre ring soliton laser
  exploiting nonlinear polarisation rotation,}
  {\protect\JournalTitle{Electronics Letters}} \textbf{28}, 1391--1393 (1992).

\bibitem{173059}
K.~{Tamura}, H.~A. {Haus}, and E.~P. {Ippen}, \enquote{Self-starting additive
  pulse mode-locked erbium fibre ring laser,}
  {\protect\JournalTitle{Electronics Letters}} \textbf{28}, 2226--2228 (1992).

\bibitem{Doran:88}
N.~J. Doran and D.~Wood, \enquote{Nonlinear-optical loop mirror,}
  {\protect\JournalTitle{Opt. Lett.}} \textbf{13}, 56--58 (1988).

\bibitem{272080}
I.~N. {Duling}, C.~. {Chen}, P.~K.~A. {Wai}, and C.~R. {Menyuk},
  \enquote{Operation of a nonlinear loop mirror in a laser cavity,}
  {\protect\JournalTitle{IEEE Journal of Quantum Electronics}} \textbf{30},
  194--199 (1994).

\bibitem{Fermann:90}
M.~E. Fermann, F.~Haberl, M.~Hofer, and H.~Hochreiter, \enquote{Nonlinear
  amplifying loop mirror,} {\protect\JournalTitle{Opt. Lett.}} \textbf{15},
  752--754 (1990).

\bibitem{Zhang:13}
Z.~Zhang, C.~Mou, Z.~Yan, K.~Zhou, L.~Zhang, and S.~Turitsyn, \enquote{Sub-100
  fs mode-locked erbium-doped fiber laser using a 45{\textdegree}-tilted fiber
  grating,} {\protect\JournalTitle{Opt. Express}} \textbf{21}, 28297--28303
  (2013).

\bibitem{Wu2010}
X.~Wu, D.~Y. Tang, L.~M. Zhao, and H.~Zhang, \enquote{Mode-locking of fiber
  lasers induced by residual polarization dependent loss of cavity components,}
  {\protect\JournalTitle{Laser Physics}} \textbf{20}, 1913--1917 (2010).

\bibitem{Lin_2014}
S.-F. Lin, H.-Y. Wang, Y.-C. Su, Y.-C. Chi, and G.-R. Lin, \enquote{Multi-order
  bunched soliton pulse generation by nonlinear polarization rotation
  mode-locking erbium-doped fiber lasers with weak or strong
  polarization-dependent loss,} {\protect\JournalTitle{Laser Physics}}
  \textbf{24}, 105113 (2014).

\bibitem{Wang:07}
Q.~Wang, G.~Rajan, P.~Wang, and G.~Farrell, \enquote{Polarization dependence of
  bend loss for a standard singlemode fiber,} {\protect\JournalTitle{Opt.
  Express}} \textbf{15}, 4909--4920 (2007).

\end{thebibliography}

\bibliographyfullrefs{sample}


\ifthenelse{\equal{\journalref}{aop}}{%
\section*{Author Biographies}
\begingroup
\setlength\intextsep{0pt}
\begin{minipage}[t][6.3cm][t]{1.0\textwidth} 
  \begin{wrapfigure}{L}{0.25\textwidth}
    \includegraphics[width=0.25\textwidth]{john_smith.eps}
  \end{wrapfigure}
  \noindent
  {\bfseries John Smith} received his BSc (Mathematics) in 2000 from The University of Maryland. His research interests include lasers and optics.
\end{minipage}
\begin{minipage}{1.0\textwidth}
  \begin{wrapfigure}{L}{0.25\textwidth}
    \includegraphics[width=0.25\textwidth]{alice_smith.eps}
  \end{wrapfigure}
  \noindent
  {\bfseries Alice Smith} also received her BSc (Mathematics) in 2000 from The University of Maryland. Her research interests also include lasers and optics.
\end{minipage}
\endgroup
}{}

\end{document}